\documentclass[12pt]{article}
\usepackage{amssym}
\def\Z{\mbox{$\Bbb Z$}}
\def\C{\mbox{$\Bbb C$}}
\def\case#1#2{{\textstyle{#1\over #2}}}
\def\ap{a^{\dagger}}
\def\bbeta{\bar{\beta}}
\def\bp{b^{\dagger}}

\title{
Spectrum generating algebra and coherent states of the $C_{\lambda}$-extended
oscillator}

\author{C.\ Quesne \thanks{Directeur de recherches FNRS; E-mail:
cquesne@ulb.ac.be} \\
{\small \sl Physique Nucl\'eaire Th\'eorique et Physique
Math\'ematique,}\\  {\small \sl Universit\'e Libre de Bruxelles, Campus de
la Plaine
CP229,}\\ {\small \sl Boulevard~du Triomphe, B-1050 Brussels, Belgium}}
\date{ }
\begin{document}
\maketitle

\begin{abstract}
$C_{\lambda}$-extended oscillator algebras, generalizing the
Calogero-Vasiliev algebra, where $C_{\lambda}$ is the cyclic group of order
$\lambda$, have recently proved very useful in the context of supersymmetric
quantum mechanics and some of its variants. Here we determine the spectrum
generating algebra of the $C_{\lambda}$-extended oscillator. We then construct
its coherent states, study their nonclassical properties, and compare the latter
with those of standard $\lambda$-photon coherent states, which are obtained as a
special case. Finally, we briefly review some other types of coherent states
associated with the $C_{\lambda}$-extended oscillator.
\end{abstract}
%
%
\section{Introduction}

Coherent states~(CS) of the harmonic oscillator~\cite{glauber} are known to
have properties similar to those of the classical radiation field.  They may be
defined in various ways, for instance as eigenstates of the oscillator
annihilation
operator $b$. With the corresponding creation operator $\bp$ and the number
operator $N_b \equiv \bp b$, the latter satisfies the commutation relations
\begin{equation}
  \left[N_b, \bp\right] = \bp, \qquad \left[N_b, b\right] = - b, \qquad
  \left[b, \bp\right] = I.  \label{eq:can-alg}
\end{equation}
\par
%
%
In contrast, generalized CS associated with various
algebras~\cite{perelomov} may
have some nonclassical properties, such as photon antibunching or sub-Poissonian
photon statistics, and squeezing. As examples of such CS, we may quote the
eigenstates of $b^2$, which were introduced as even and odd CS or cat
states~\cite{dodonov}, and are a special case of generalized CS associated
with the
Lie algebra su(1,1)~\cite{barut}. We may also mention the eigenstates of
$b^{\lambda}$ ($\lambda>2$) or kitten states~\cite {buzek}, which may be
generated in $\lambda$-photon processes.\par
%
%
Other examples are provided by nonlinear CS associated with a deformed
oscillator
(or $f$-oscillator). The latter is defined in terms of creation,
annihilation, and
number operators, $\ap = f(N_b) \bp$, $a = b f(N_b)$, $N = N_b$, satisfying the
commutation relations~\cite{daska,solomon}
\begin{equation}
  \left[N, \ap\right] = \ap, \qquad [N, a] = - a, \qquad \left[a,
\ap\right] = G(N),
  \label{eq:def-alg}
\end{equation}
where $f$ is some Hermitian operator-valued function of the number operator and
$G(N) =  (N+1) f^2(N+1) - N f^2(N)$. Nonlinear CS, defined as eigenstates of
$a$~\cite{solomon,matos,manko}, of $a^2$~\cite{mancini}, or of an arbitrary
power
$a^{\lambda}$ ($\lambda > 2$)~\cite{liu}, have been considered in
connection with
nonclassical properties. It has been shown that for a particular class of
nonlinearities the first ones are useful in the description of a trapped
ion~\cite{matos}.\par
%
%
In the present communication, we shall consider some multiphoton CS, which
may be
associated with the recently introduced $C_{\lambda}$-extended
oscillator~\cite{cq98}. The latter may be considered as a deformed
oscillator with
a \Z$_{\lambda}$-graded Fock space and has proved very useful in the context of
supersymmetric quantum mechanics and some of its variants~\cite{cq98,cq99}. In
particular, we shall deal here in detail with CS of the $C_{\lambda}$-extended
oscillator spectrum generating algebra~\cite{cq00}, which are a special
case of the
CS of Ref.~\cite{liu} and exhibit some nonclassical properties.\par
%
%
\section{\boldmath The $C_{\lambda}$-extended oscillator algebra}

The $C_{\lambda}$-extended oscillator algebra (where $C_{\lambda} =
\Z_{\lambda}$ is the cyclic group of order~$\lambda$) was introduced as a
generalization of the Calogero-Vasiliev algebra, defined by
\begin{equation}
  \left[N, \ap\right] = \ap, \qquad \left[a, \ap\right] = I + \alpha_0 K, \qquad
  \left\{K, \ap\right\} = 0,  \label{eq:Calogero}
\end{equation}
and their Hermitian conjugates, where $\alpha_0$ is some real parameter subject
to the condition $\alpha_0 > -1$, and $K$ is some Hermitian operator. The latter
may be realized as $K = (-1)^N$, so that the second equation
in~(\ref{eq:Calogero})
becomes equivalent to $\left[a, \ap\right] = I + \alpha_0 P_0 + \alpha_1 P_1$,
where $\alpha_0 + \alpha_1 = 0$ and $P_0 = \frac{1}{2} \left[I + (-1)^N\right]$,
$P_1 = \frac{1}{2} \left[I - (-1)^N\right]$ project on the even and odd
subspaces of
the Fock space $\cal F$, respectively.\par
%
%
When partitioning $\cal F$ into $\lambda$ subspaces ${\cal F}_{\mu} \equiv
\{\, |k\lambda + \mu\rangle \mid k = 0, 1, \ldots\,\}$, $\mu=0$, 1,
\ldots,~$\lambda-1$, instead of two, the Calogero-Vasiliev algebra is
replaced by
the $C_{\lambda}$-extended oscillator algebra, defined by~\cite{cq98}
\begin{equation}
  \left[N, \ap\right] = \ap, \qquad \left[a, \ap\right] = I +
\sum_{\mu=0}^{\lambda-1}
  \alpha_{\mu} P_{\mu}, \qquad \ap P_{\mu} = P_{\mu+1} \ap, \label{eq:alg-def}
\end{equation}
and their Hermitian conjugates, where $P_{\mu} = \lambda^{-1}
\sum_{\nu=0}^{\lambda-1} \exp[2\pi {\rm i} \nu (N-\mu)/\lambda]$ projects on
${\cal F}_{\mu}$, $\sum_{\mu=0}^{\lambda-1} P_{\mu} = I$, and $\alpha_{\mu}$ are
some real parameters subject to the conditions $\sum_{\mu=0}^{\lambda-1}
\alpha_{\mu} = 0$ and $\sum_{\nu=0}^{\mu-1} \alpha_{\nu} > - \mu$, $\mu = 1$, 2,
\ldots,~$\lambda-1$. Taking this form of $P_{\mu}$ into account, it is
clear that the
$C_{\lambda}$-extended oscillator algebra~(\ref{eq:alg-def}) is a special case
of deformed oscillator algebra, as defined in~(\ref{eq:def-alg}).\par
%
%
The operators $N$, $\ap$, $a$ are related to each other through the structure
function $F(N) = N + \sum_{\mu=0}^{\lambda-1} \beta_{\mu} P_{\mu}$,
$\beta_{\mu} \equiv \sum_{\nu=0}^{\mu-1} \alpha_{\nu}$, which is a fundamental
concept of deformed oscillators: $\ap a = F(N)$, $a \ap = F(N+1)$~\cite{daska,
solomon}. Comparing with Eq.~(\ref{eq:def-alg}), we get
$f(N) = (F(N)/N)^{1/2}$.\par
%
%
The Fock space basis states $|n\rangle = |k \lambda + \mu\rangle = {\cal
N}_n^{-1/2}
\left(\ap\right)^n |0\rangle$, where $a |0\rangle = 0$, $k = 0$, 1,~\ldots, and
$\mu=0$, 1,~\ldots, $\lambda-1$, satisfy the relations
\begin{equation}
  N |n\rangle = n |n\rangle, \qquad \ap |n\rangle = \sqrt{F(n+1)}\,
|n+1\rangle, \qquad
  a |n\rangle = \sqrt{F(n)}\, |n-1\rangle. \label{eq:op-action}
\end{equation}
Due to the restrictions on the range of the parameters $\alpha_{\mu}$ given
below
Eq.~(\ref{eq:alg-def}), $F(\mu) = \beta_{\mu} + \mu > 0$ so that all the states
$|n\rangle$ are well defined.\par
%
%
The $C_{\lambda}$-extended oscillator Hamiltonian is defined by~\cite{cq98}
\begin{equation}
  H_0 = \case{1}{2} \left\{a, \ap\right\}.  \label{eq:hamiltonian}
\end{equation}
Its eigenstates are the states $|n\rangle = |k\lambda + \mu\rangle$ and
their eigenvalues are given by $E_{k\lambda+\mu} = k\lambda + \mu +
\gamma_{\mu} + \frac{1}{2}$, where $\gamma_{\mu} \equiv \frac{1}{2}
\left(\beta_{\mu} + \beta_{\mu+1}\right)$. In each ${\cal F}_{\mu}$ subspace of
$\cal F$, the spectrum of $H_0$ is harmonic, but the $\lambda$ infinite sets of
equally spaced energy levels, corresponding to $\mu=0$, 1,~\ldots, $\lambda-1$,
are shifted with respect to each other by some amounts depending upon the
parameters $\alpha_0$, $\alpha_1$,~\ldots, $\alpha_{\lambda-1}$.\par
%
%
\section{\boldmath Spectrum generating algebra of the $C_{\lambda}$-extended
oscillator}

One can generate the whole spectrum of the $C_{\lambda}$-extended oscillator
Hamiltonian~(\ref{eq:hamiltonian}) from the eigenstates $|\mu\rangle$, $\mu=0$,
1, \ldots,~$\lambda-1$, by using the operators~\cite{cq00}
\begin{equation}
  J_+ = \frac{1}{\lambda} \left(\ap\right)^{\lambda}, \qquad J_- =
  \frac{1}{\lambda} a^{\lambda}, \qquad J_0 = \frac{1}{\lambda} H_0 =
  \frac{1}{2\lambda} \left\{a, \ap\right\}. \label{eq:defsu-gen}
\end{equation}
They satisfy the commutation relations
\begin{equation}
  [J_0, J_{\pm}] = \pm J_{\pm}, \qquad [J_+, J_-] = f(J_0, P_{\mu}), \qquad
[J_0,
  P_{\mu}] = [J_{\pm}, P_{\mu}] = 0 ,  \label{eq:defsu-com}
\end{equation}
where $f(J_0, P_{\mu})$ (which has nothing to do with the function $f(N)$ of
Eq.~(\ref{eq:def-alg})) is a ($\lambda-1$)th-degree polynomial in $J_0$ with
$P_{\mu}$-dependent coefficients, $f(J_0, P_{\mu}) = \sum_{i=0}^{\lambda-1}
s_i(P_{\mu}) J_0^i$. The spectrum generating algebra~(SGA) of the
$C_{\lambda}$-extended oscillator is therefore a $C_{\lambda}$-extended
polynomial deformation of su(1,1): in each ${\cal F}_{\mu}$ subspace, it
reduces to
a standard polynomial deformation of su(1,1)~\cite{poly}.\par
%
%
Its Casimir operator can be written as
\begin{equation}
  C = J_- J_+ + h(J_0, P_{\mu}) = J_+ J_- + h(J_0, P_{\mu}) - f(J_0, P_{\mu}),
  \label{eq:defsu-Cas}
\end{equation}
where $h(J_0, P_{\mu})$ is a $\lambda$th-degree polynomial in $J_0$ with
$P_{\mu}$-dependent coefficients, $h(J_0, P_{\mu}) = \sum_{i=0}^{\lambda}
t_i(P_{\mu}) J_0^i$. Each ${\cal F}_{\mu}$ subspace is the carrier space of a
unitary irreducible representation (unirrep) of the SGA, characterized by an
eigenvalue $c_{\mu}$ of $C$, and by the lowest eigenvalue $\left(\mu +
\gamma_{\mu} + \case{1}{2}\right)/{\lambda}$ of $J_0$. The explicit
expressions of $f(J_0, P_{\mu})$, $h(J_0,P_{\mu})$, and $c_{\mu}$ are given in
Ref.~\cite{cq00}.\par
%
%
{}For $\lambda=2$, for which the $C_{\lambda}$-extended oscillator algebra
reduces to the Calogero-Vasiliev algebra, the SGA (\ref{eq:defsu-gen}),
(\ref{eq:defsu-com}) reduces to the Lie algebra su(1,1), for which $f(J_0) = -2
J_0$, $h(J_0) = - J_0 (J_0+1)$, and $c = (1 + \alpha_{\mu})(3 - \alpha_{\mu})/16
$~\cite{brze}.\par
%
%
Nonlinearities make  their appearance for $\lambda=3$, for which
\begin{eqnarray}
  f(J_0, P_{\mu}) & = & - 9 J_0^2 - J_0 \sum_{\mu} (\alpha_{\mu} +
        2\alpha_{\mu+1}) P_{\mu} - \case{1}{12} \sum_{\mu} (1 + \alpha_{\mu})
        (5 - \alpha_{\mu}) P_{\mu}, \nonumber \\
  h(J_0, P_{\mu}) & = & - J_0 \biggl[3 J_0^2 + \case{1}{2} J_0 \sum_{\mu}
        (9 + \alpha_{\mu} + 2\alpha_{\mu+1}) P_{\mu} + \case{1}{12} \sum_{\mu}
        \bigl(23 + 10 \alpha_{\mu} \nonumber \\
  && \mbox{} + 12 \alpha_{\mu+1} - \alpha_{\mu}^2\bigr) P_{\mu}\biggr],
        \nonumber  \\
  c_{\mu} & = & \case{1}{72} (1 + \alpha_{\mu}) (5 - \alpha_{\mu}) (3 +
        \alpha_{\mu} + 2 \alpha_{\mu+1}).
\end{eqnarray}
\par
%
%
{}For $\alpha_{\mu} = 0$ corresponding to $\ap = \bp$, $a = b$, the
operators~(\ref{eq:defsu-gen}) close a polynomial deformation of su(1,1), with
$f(J_0)$ and $h(J_0)$ expressed in terms of some binomial coefficients and
Stirling
numbers~\cite{cq00}.\par
%
%
\section{\boldmath Coherent states associated with the $C_{\lambda}$-extended
oscillator spectrum generating algebra}

As CS associated with the $C_{\lambda}$-extended oscillator SGA, let us
consider
generalizations of the Barut-Girardelle CS of su(1,1)~\cite{barut}, to
which they
will reduce in the case $\lambda=2$. These are  the eigenstates $|z;
\mu\rangle$ of
the operator $J_-$ defined in~(\ref{eq:defsu-gen}),
\begin{equation}
  J_- |z; \mu\rangle = z |z; \mu\rangle, \qquad z \in \C, \qquad \mu=0, 1,
\ldots,
  \lambda-1.  \label{eq:CS-def}
\end{equation}
Here $\mu$ distinguishes between the $\lambda$ independent (and orthogonal)
solutions of equation~(\ref{eq:CS-def}), belonging to the various subspaces
${\cal
F}_{\mu}$. The CS $|z; \mu\rangle$ may be considered as special cases of the
nonlinear CS of Ref.~\cite{liu}, since Eq.~(\ref{eq:CS-def}) is equivalent to
$a^{\lambda} |z; \mu\rangle = \lambda z |z; \mu\rangle$, for $a = b f(N_b)$ and
$f(N_b)$ as given in Sec.~2.\par
%
%
It can be shown~\cite{cq00} that the states~(\ref{eq:CS-def}) satisfy Klauder's
minimal set of conditions for generalized CS~\cite{klauder}: they are
normalizable,
continuous in the label $z$, and they allow a resolution of unity. The other
discrete label $\mu$ is analogous to the vector components of vector (or
partially)
CS~\cite{deenen}.\par
%
%
The states $|z; \mu\rangle$ can be written in either of the alternative forms
\begin{equation}
  |z; \mu\rangle = \left[N_{\mu}(|z|)\right]^{-1/2} \sum_{k=0}^{\infty}
  \frac{\left(z/\lambda^{(\lambda-2)/2}\right)^k}{\left[k!\,
  \left(\prod_{\nu=1}^{\mu} (\bbeta_{\nu}+1)_k\right)
  \left(\prod_{\nu'=\mu+1}^{\lambda-1} (\bbeta_{\nu'})_k\right)\right]^{1/2}}
  |k \lambda + \mu\rangle,
\end{equation}
\begin{equation}
  |z; \mu\rangle =\left[N_{\mu}(|z|)\right]^{-1/2} {}_0F_{\lambda-1}
  \left(\bbeta_1+1, \ldots, \bbeta_{\mu}+1, \bbeta_{\mu+1}, \ldots,
  \bbeta_{\lambda-1}; z J_+/\lambda^{\lambda-2}\right) |\mu\rangle,
  \label{eq:CS-exp}
\end{equation}
where $\bbeta_{\mu} \equiv (\beta_{\mu} + \mu)/\lambda$, $(a)_k$ denotes
Pochhammer's symbol, and the normalization factor $N_{\mu}(|z|)$ can be
expressed
in terms of a generalized hypergeometric function,
\begin{equation}
  N_{\mu}(|z|) = {}_0F_{\lambda-1} \left(\bbeta_1+1, \ldots, \bbeta_{\mu}+1,
  \bbeta_{\mu+1}, \ldots, \bbeta_{\lambda-1}; y\right), \qquad y \equiv
  |z|^2/\lambda^{\lambda-2}.  \label{eq:normalization}
\end{equation}
\par
%
%
Their unity resolution relation can be written as
\begin{equation}
  \sum_{\mu} \int d\rho_{\mu}\left(z, z^*\right) |z; \mu\rangle \langle z;
\mu| = I,
  \label{eq:unity}
\end{equation}
where $d\rho_{\mu}\left(z, z^*\right)$ is a positive measure, given in
terms of a
generalized hypergeometric function and a Meijer $G$-function by
\begin{eqnarray}
  d\rho_{\mu}\left(z, z^*\right) & = & {}_0F_{\lambda-1} \left(\bbeta_1+1,
\ldots,
         \bbeta_{\mu}+1, \bbeta_{\mu+1}, \ldots, \bbeta_{\lambda-1}; y\right)
         h_{\mu}(y) |z| d|z| d\phi, \nonumber \\
  h_{\mu}(y) & = & \frac{G^{\lambda 0}_{0 \lambda} \left(y \mid 0,
\bbeta_1, \ldots,
         \bbeta_{\mu}, \bbeta_{\mu+1}-1, \ldots, \bbeta_{\lambda-1}-1\right)}
        {\pi \lambda^{\lambda-2} \left(\prod_{\nu=1}^{\mu} \Gamma
        (\bbeta_{\nu}+1)\right) \left(\prod_{\nu'=\mu+1}^{\lambda-1} \Gamma
        (\bbeta_{\nu'})\right)},  \label{eq:measure}
\end{eqnarray}
with $y$ defined in Eq.~(\ref{eq:normalization}).\par
%
%
In the $\lambda = 2$ case,  the functions ${}_0F_1$ and $G^{20}_{02}$ of
Eqs.~(\ref{eq:CS-exp}), (\ref{eq:normalization}), and~(\ref{eq:measure}) being
proportional to modified Bessel functions $I_{2\nu}(2|z|)$ and $K_{2\nu}(2|z|)$,
$\nu = (\alpha_0 - 1 + 2\mu)/2$, respectively, the CS defined
in~(\ref{eq:CS-def})
reduce to Barut-Girardello su(1,1) CS~\cite{barut} for the appropriate unirreps,
as it should be.\par
%
%
{}For $\alpha_{\mu} = 0$ corresponding to $\ap = \bp$, $a = b$, the
CS defined in~(\ref{eq:CS-def}) reduce to the eigenstates of $b^{\lambda}$ or
standard $\lambda$-photon CS~\cite{buzek},
\begin{equation}
  |z; \mu\rangle = [N_{\mu}(|z|)]^{-1/2} \sum_{k=0}^{\infty}
  \left(\frac{\mu!}{(k\lambda+\mu)!}\right)^{1/2} (\lambda z)^k |k\lambda
  + \mu\rangle,  \label{eq:buzek-CS}
\end{equation}
satisfying the resolution of unity~(\ref{eq:unity}) with $h_{\mu}(y)$ given by
\begin{equation}
  h_{\mu}(y) = \lambda^{\mu-\lambda+2} \left(\pi \mu!\right)^{-1}
  y^{(\mu-\lambda+1)/\lambda} \exp\left(- \lambda y^{1/\lambda}\right).
  \label{eq:buzek-weight}
\end{equation}
The states~(\ref{eq:buzek-CS}) can be rewritten in the alternative form
\begin{equation}
  |z;\mu\rangle = \left(\frac{\mu!}{E_{\lambda,\mu+1}\left(\lambda^2
  |z|^2\right)}\right)^{1/2} E_{\lambda,\mu+1}\left(\lambda^2 z J_+\right)
  |\mu\rangle,
\end{equation}
where $E_{\alpha,\beta}(x) \equiv \sum_{k=0}^{\infty} x^k/\Gamma(\alpha k +
\beta)$ is a generalized Mittag-Leffler function. Hence, they provide a simple
example of the Mittag-Leffler CS considered in Ref.~\cite{sixdeniers}.\par
%
%
\section{Nonclassical properties of coherent states}

The CS $|z; \mu\rangle$ may be considered as exotic states in quantum optics.
Their properties may be analyzed in two different ways, by considering either
``real'' photons, described by the operators $\bp$, $b$ satisfying the canonical
commutation relation, as given in Eq.~(\ref{eq:can-alg}), or ``dressed''
photons,
described by the operators $\ap$, $a$ of Eq.~(\ref{eq:def-alg}), which may
appear in
some phenomenological models explaining some non-intuitive observable
phenomena.\par
%
%
\subsection{Photon statistics}

Since $N = N_b$, the photon number statistics is not affected by the choice
made for
the type of photons. A measure of its deviation from the Poisson
distribution is the
Mandel parameter
\begin{equation}
  Q = \frac{\langle(\Delta N)^2\rangle - \langle N \rangle}{\langle N \rangle},
  \qquad \Delta N \equiv N - \langle N \rangle,
\end{equation}
which vanishes for the Poisson distribution, and is positive or negative
according to
whether the distribution is super-Poissonian (bunching effect) or sub-Poissonian
(antibunching effect).\par
%
%
It is well known that for $\lambda=2$, the standard even (resp.\ odd) CS,
corresponding to $\alpha_0 = \alpha_1 = 0$ or $\ap = \bp$, $a = b$ and $\mu = 0$
(resp.\ $\mu = 1$), are characterized by a super-Poissonian (resp.\
sub-Poissonian)
number distribution. It can be shown~\cite{cq00} that for the even (resp.\
odd) CS
associated with the Calogero-Vasiliev algebra, i.e., for $\lambda = 2$,
$\alpha_0 =
- \alpha_1 \ne 0$ and $\mu = 0$ (resp.\ $\mu = 1$), this trend is enhanced for
positive (resp.\ negative) values of $\alpha_0$. However, as shown in
Fig.~1, for
negative (resp.\ positive) values of $\alpha_0$ and sufficiently high
values of $|z|$,
the opposite trend can be seen.\par
%
%
{}For higher values of $\lambda$, more or less similar results are obtained for
$\mu=0$, on one hand, and $\mu \ne 0$, on the other hand. However the
behaviour of $Q$ becomes more complicated for intermediate values of
$\mu$~\cite{cq00}.\par
%
%
\subsection{Squeezing effect}

\subsubsection{``Dressed'' photons}

Let us define the deformed quadratures $x$ and $p$ as
\begin{equation}
  x = \frac{1}{\sqrt{2}} \left(\ap + a\right), \qquad p = \frac{{\rm
i}}{\sqrt{2}}
  \left(\ap - a\right).
\end{equation}
In any state belonging to ${\cal F}_{\mu}$, their dispersions $\langle
(\Delta x)^2
\rangle$ and $\langle (\Delta p)^2 \rangle$ satisfy the uncertainty relation
\begin{equation}
 \langle (\Delta x)^2 \rangle \langle (\Delta p)^2 \rangle \ge \frac{1}{4}
|\langle [x,
  p]\rangle|^2 = \frac{\lambda^2}{4} (\bbeta_{\mu+1} - \bbeta_{\mu})^2,
  \label{eq:UR}
\end{equation}
where the right-hand side becomes smaller than the conventional value 1/4 if
$\alpha_0 < 0$ for $\mu=0$ or $-2 < \alpha_{\mu} < 0$ for $\mu=1$, 2, \ldots, or
$\lambda-1$.\par
%
%
In ${\cal F}_{\mu}$, the role of the vacuum state is played by the number state
$|\mu\rangle = |0;\mu\rangle$, which is annihilated by $J_-$. The corresponding
dispersions are given by
\begin{equation}
  \langle (\Delta x)^2\rangle_0 = \langle (\Delta p)^2\rangle_0 =
\frac{\lambda}{2}
  (\bbeta_{\mu+1} + \bbeta_{\mu}).
  \label{eq:dispersions-vac}
\end{equation}
Comparing with the uncertainty relation~(\ref{eq:UR}), we conclude that the
state
$|\mu\rangle$ satisfies the minimum uncertainty property in ${\cal
F}_{\mu}$, i.e.,
gives rise to the equality in~(\ref{eq:UR}), only for $\mu=0$ because
$\bbeta_0 =
0$ and $\bbeta_{\mu} > 0$ for $\mu=1$, 2,~\ldots, $\lambda-1$. On the other
hand,
the dispersions in the vacuum may be smaller than the conventional value 1/2 for
$\mu=0$, 1,~\ldots, $\lambda-2$.\par
%
%
Let us restrict ourselves to the CS $|z; 0\rangle$, which satisfies for $z=0$
the minimum uncertainty property. The quadrature $x$ (resp.\ $p$) is said to be
squeezed to the second order in $|z;0\rangle$ if $X \equiv \langle (\Delta
x)^2\rangle/\langle (\Delta x)^2\rangle_0$ (resp.\ $P \equiv \langle (\Delta
p)^2\rangle/\langle (\Delta p)^2\rangle_0$) is less than one. Similarly, it
is said to
be squeezed to the fourth order if $Y \equiv \langle (\Delta x)^4\rangle/\langle
(\Delta x)^4\rangle_0$ (resp.\ $Q \equiv \langle (\Delta
p)^4\rangle/\langle (\Delta
p)^4\rangle_0$) is less than one. \par
%
%
{}For $\lambda = 2$, $X$ and $P$, or $Y$ and $Q$, are related with each other by
the transformation ${\rm Re} z \to - {\rm Re} z$. Moreover $X$ and $Y$ are
minimum
for real, negative values of $z$. On Fig.~2, they are displayed for such
values. We
note a large squeezing effect over the whole range of real, negative values
of $z$
for positive values of $\alpha_0$ (for which the conventional uncertainty
relation
is respected).\par
%
%
{}For $\lambda > 2$, there is no second-order squeezing, but for $\lambda = 4$, a
small fourth-order squeezing is obtained in accordance with the results for
standard $\lambda$-photon CS~\cite{buzek}.\par
%
%
\subsubsection{``Real'' photons}

Let us now define the quadratures $x$ and $p$ as
\begin{equation}
  x = \frac{1}{\sqrt{2}} \left(\bp + b\right), \qquad p = \frac{{\rm
i}}{\sqrt{2}}
  \left(\bp - b\right).
\end{equation}
Their dispersions $\langle (\Delta x)^2 \rangle$ and $\langle (\Delta p)^2
\rangle$
satisfy the usual uncertainty relation. Considering again the CS $|z;
0\rangle$, on
Fig.~3 we observe for the ratios $X$ and $P$  more or less similar trends
as noted in
the case of ``dressed'' photons.\par
%
%
\section{Concluding remarks}

In the present contribution, we determined the SGA of the $C_{\lambda}$-extended
oscillator and studied some CS associated with it, namely the eigenstates of its
lowering generator~$J_-$.\par
%
%
Other types of CS may be considered and will be studied in a forthcoming
publication. Let us mention here two of them:

\begin{enumerate}

\item The eigenstates of the $C_{\lambda}$-extended oscillator
annihilation operator~$a$:
\begin{equation}
  a |z; \mu\rangle = z |z; \mu\rangle.
\end{equation}
These generalize the paraboson CS, which correspond to $\lambda =
2$~\cite{sharma}.

\item The solutions of the equation
\begin{equation}
  \left[a^{\lambda - \alpha} - z \left(\ap\right)^{\alpha}\right] |z;
\mu\rangle = 0,
  \quad \alpha = 0, 1, \ldots, \left[\frac{\lambda}{2}\right], \quad \mu = 0, 1,
  \ldots, \lambda - \alpha - 1.
\end{equation}
For $\alpha=0$, these are the eigenstates of~$a^{\lambda}$, which are directly
related to those of~$J_-$, considered here. Moreover, for $\lambda=2$ and
$\alpha=1$, they reduce to the Perelomov su(1,1) CS~\cite{perelomov}.

\end{enumerate}
%
%

\end{document}